
\magnification \magstep1
\raggedbottom
\openup 4\jot
\voffset6truemm
\def\cstok#1{\leavevmode\thinspace\hbox{\vrule\vtop{\vbox{\hrule\kern1pt
\hbox{\vphantom{\tt/}\thinspace{\tt#1}\thinspace}}
\kern1pt\hrule}\vrule}\thinspace}
\headline={\ifnum\pageno=1\hfill\else
\hfill{\it $\alpha$-Surfaces for Complex Space-Times with Torsion}
\hfill \fi}
\rightline {November 1991, DSF-91/20}
\centerline {\bf $\alpha$-SURFACES FOR COMPLEX SPACE-TIMES}
\centerline {\bf WITH TORSION}
\vskip 1cm
\centerline {\bf Giampiero Esposito}
\vskip 1cm
\leftline {\it Istituto Nazionale di Fisica Nucleare,
Sezione di Napoli, Gruppo IV,}
\leftline {\it Mostra d'Oltremare, Padiglione 20, 80125 Napoli}
\leftline {\it Dipartimento di Scienze Fisiche,}
\leftline {\it Mostra d'Oltremare, Padiglione 19, 80125 Napoli}
\vskip 1cm
\noindent
{\bf Abstract. -}
This paper studies necessary conditions for the
existence of $\alpha$-surfaces in complex space-time manifolds
with nonvanishing torsion. For these manifolds,
Lie brackets of vector fields and spinor Ricci
identities contain explicitly the effects of torsion. This
leads to an integrability condition for
$\alpha$-surfaces which does not
involve just the self-dual Weyl spinor, as in
complexified general relativity, but
also the torsion spinor, in a nonlinear way, and its covariant
derivative. Interestingly,
a particular solution of the integrability condition is given
by right-flat and right-torsion-free space-times.
\vskip 1cm
\leftline {PACS 04.20.Cv - Fundamental problems and general formalism.}
\leftline {PACS 04.50 - Unified field theories and other theories
of gravitation.}
\vskip 100cm
Twistor theory was created by Roger Penrose as an
approach to the quantum-gravity problem where null surfaces
and some particular complex manifolds are regarded as
fundamental entities, whereas space-time points are only derived
objects, since they might become ill-defined at the Planck length
[1]. It is by now well-known that the building blocks of classical
field theory in Minkowski space-time are the $\alpha$-planes. In other
words, one first takes complexified compactified Minkowski space
$CM^{\#}$, and one then defines $\alpha$-planes as null two-surfaces,
such that the metric vanishes over them, and their null tangent vectors
have the two-component-spinor form $\lambda^{A}\pi^{A'}$, where
$\lambda^{A}$ is varying and $\pi^{A'}$ is fixed by a well-known
differential equation [1]. This definition can be generalized to complex
or real Riemannian (i.e. with positive-definite metric) space-times
provided the Weyl curvature is anti-self-dual, giving rise to the
so-called $\alpha$-surfaces [1].

We are here interested in the case of complex space-times with
nonvanishing torsion (hereafter referred to as $CU_{4}$ space-times).
This study appears relevant at least for the following reasons:
(i) the definition of torsion is a peculiarity of relativistic theories
of gravitation; (ii) the gauge theory of the Poincar\'e group leads
to theories with torsion; (iii) theories with torsion are theories of
gravity with second-class constraints; (iv) if torsion is nonvanishing,
the occurrence of cosmological singularities can be less generic than in
general relativity [2]. We have thus studied the problem: what are the
conditions on curvature and torsion for a $CU_{4}$ space-time to
admit $\alpha$-surfaces (defined as above) ?

The starting point of our calculation, based on the use of the full
$U_{4}$-connection with torsion, is the evaluation of the Lie
bracket of two vector fields $X$ and $Y$ tangent to a totally null
two-surface in $CU_{4}$. By virtue of Frobenius' theorem [1], this Lie
bracket is a linear combination of $X$ and $Y$
$$
[X,Y]=\varphi X + \rho Y
\; \; \; \; ,
\eqno (1)
$$
where $\varphi$ and $\rho$ are scalar functions. Moreover, using the
definition of the torsion tensor $S(X,Y)$, one also has
$$
[X,Y] \equiv \nabla_{X}Y -\nabla_{Y}X -2S(X,Y)
\; \; \; \; .
\eqno (2)
$$
Thus, since in $CU_{4}$ models the torsion tensor,
antisymmetric in the first two indices, can be expressed spinorially
as
$$
S_{ab}^{\; \; \; c}=\chi_{AB}^{\; \; \; \; \; CC'}\epsilon_{A'B'}+
{\widetilde \chi}_{A'B'}^{\; \; \; \; \; \; \; CC'}
\epsilon_{AB}
\; \; \; \; ,
\eqno (3)
$$
where the spinors $\chi$ and ${\widetilde \chi}$ are symmetric
in $AB$ and $A'B'$ respectively, and are {\it totally independent},
one finds by comparison of Eqs. (1) and (2) that
$$
\pi^{A'}\Bigr(\nabla_{AA'}\pi_{B'}\Bigr)=\xi_{A}\pi_{B'}
-2\pi^{A'}\pi^{C'}{\widetilde \chi}_{A'B'AC'}
\; \; \; \; ,
\eqno (4)
$$
where we have set $X^{a}=\lambda^{A}\pi^{A'}$, $Y^{a}=\mu^{A}\pi^{A'}$,
while $\varphi$ and $\rho$ are obtained as follows:
$-\mu^{A}\xi_{A}=\varphi$, $\lambda^{A}\xi_{A}=\rho$.
This is the desired necessary condition for the field
$\pi^{A'}$ to define an $\alpha$-surface in the presence of torsion.
We now have to derive the integrability condition for Eq. (4). For this
purpose, we operate with $\pi^{B'}\pi^{C'}\nabla_{\; \; C'}^{A}$ on
both sides of Eq. (4), we repeatedly use
the Leibniz rule and Eq. (4), and we
also use the spinor formula for the Riemann tensor and spinor Ricci
identities. The relations we need are
$$ \eqalignno{
R_{abcd}&=\psi_{ABCD}\epsilon_{A'B'}\epsilon_{C'D'}
+{\widetilde \psi}_{A'B'C'D'}\epsilon_{AB}\epsilon_{CD}\cr
&+\Phi_{ABC'D'}\epsilon_{A'B'}\epsilon_{CD}
+{\widetilde \Phi}_{A'B'CD}\epsilon_{AB}\epsilon_{C'D'}\cr
&+\Sigma_{AB}\epsilon_{A'B'}\epsilon_{CD}\epsilon_{C'D'}
+{\widetilde \Sigma}_{A'B'}\epsilon_{AB}\epsilon_{CD}
\epsilon_{C'D'}\cr
&+\Lambda \Bigr(\epsilon_{AC}\epsilon_{BD}+\epsilon_{AD}
\epsilon_{BC}\Bigr)\epsilon_{A'B'}\epsilon_{C'D'}\cr
&+{\widetilde \Lambda}\Bigr(\epsilon_{A'C'}
\epsilon_{B'D'}+\epsilon_{A'D'}\epsilon_{B'C'}\Bigr)
\epsilon_{AB}\epsilon_{CD}
\; \; \; \; ,
&(5)\cr}
$$
$$
\left[\nabla_{C(A'}\nabla_{\; \; B')}^{C}
-2{\widetilde \chi}_{A'B'}^{\; \; \; \; \; \; \; HH'}
\nabla_{HH'}\right]\pi^{C'}
={\widetilde \psi}_{A'B'E'}^{\; \; \; \; \; \; \; \; \; \; \; C'}
\; \pi^{E'}
-2{\widetilde \Lambda}\pi_{(A'}\epsilon_{B')}^{\; \; \; \; C'}
+{\widetilde \Sigma}_{A'B'}\pi^{C'}
\; \; \; \; .
\eqno (6)
$$
As usual, the {\it twiddle} symbol denotes spinor or scalar quantities
independent of their untwiddled counterpart. The spinors $\psi$ and
${\widetilde \psi}$ are the anti-self-dual and self-dual
Weyl spinors respectively, and are thus invariant under
conformal rescalings of the metric. The symmetric spinors $\Sigma$ and
${\widetilde \Sigma}$ express the antisymmetric part of the Ricci
tensor: $R_{[ab]}=\Sigma_{AB}\epsilon_{A'B'}+{\widetilde \Sigma}_{A'B'}
\epsilon_{AB}$. More details on the spinors $\Phi$, ${\widetilde \Phi}$
and the scalar functions $\Lambda$, ${\widetilde \Lambda}$ can be found
in Ref. [3]. Eq. (6) is proved defining the operator
$\cstok{\ }_{ab} \equiv 2\nabla_{[a}\nabla_{b]}-2S_{ab}^{\; \; \; c}
\; \nabla_{c}$
and the self-dual null bivector
$k^{ab} \equiv \kappa^{A}\kappa^{B}\epsilon^{A'B'}$,
and using the tensor form of the Ricci identity for $U_{4}$ theories
$$
\cstok{\ }_{ab}k^{cd}=R_{abe}^{\; \; \; \; \; c} \; k^{ed}
+R_{abe}^{\; \; \; \; \; d} \; k^{ce}
\; \; \; \; .
\eqno (7)
$$
Thus, using the identity
$2\nabla_{[a}\nabla_{b]}=\epsilon_{A'B'}\cstok{\ }_{AB}+
\epsilon_{AB}\cstok{\ }_{A'B'}$,
where
$\cstok{\ }_{C'A'} \equiv
\nabla_{A(C'}\nabla_{\; \; A')}^{A}$,
a lengthy calculation of Eq. (7)
yields the spinorial equations
$$
\Bigr[\cstok{\ }_{AB}-2\chi_{AB}^{\; \; \; \; \; HH'}\nabla_{HH'}\Bigr]
\kappa^{C}=
\psi_{ABE}^{\; \; \; \; \; \; \; \; C} \; \kappa^{E}
-2\Lambda \kappa_{(A}\epsilon_{B)}^{\; \; \; C}
+\Sigma_{AB}\kappa^{C}
\; \; \; \; ,
\eqno (8)
$$
$$
\Bigr[\cstok{\ }_{A'B'}-2{\widetilde \chi}_{A'B'}^{\; \; \; \; \; \; \; HH'}
\nabla_{HH'}\Bigr]\kappa^{C}
={\widetilde \Phi}_{A'B'E}^{\; \; \; \; \; \; \; \; \; \; \; C} \;
\kappa^{E}
+{\widetilde \Sigma}_{A'B'}\kappa^{C}
\; \; \; \; .
\eqno (9)
$$
Eq. (6) is then obtained replacing unprimed indices with primed indices,
untwiddled spinors and scalars with twiddled spinors and scalars,
and $\kappa^{C}$ with $\pi^{C'}$ in Eq. (8).

Finally, using Eq. (4), Eq. (6), the identity
$$ \eqalignno{
\pi^{B'}\pi^{C'}\nabla_{\; \; C'}^{A}
\Bigr[\pi^{A'}\Bigr(\nabla_{AA'}\pi_{B'}\Bigr)\Bigr]&=
\pi^{B'}\pi^{C'}\Bigr(\nabla_{\; \; C'}^{A}\pi^{A'}\Bigr)
\Bigr(\nabla_{AA'}\pi_{B'}\Bigr)\cr
&-\pi^{A'}\pi^{B'}\pi^{C'}
\Bigr(\cstok{\ }_{C'A'}\pi_{B'}\Bigr) \; \; \; \; ,
&(10)\cr}
$$
and the well-known property $\lambda_{A}\lambda^{A}=
\pi_{A'}\pi^{A'}=0$, the integrability condition for $\alpha$-surfaces
is found to be
$$ \eqalignno{
{\widetilde \psi}_{A'B'C'D'}&=
-4{\widetilde \chi}_{A'B'AL'}
{\widetilde \chi}_{C' \; \; \; \; \; \; D'}^{\; \; \; \; L'A}
+4{\widetilde \chi}_{L'B'AC'}
{\widetilde \chi}_{A'D'}^{\; \; \; \; \; \; \; \; AL'}\cr
&+2\nabla_{\; \; D'}^{A}\Bigr({\widetilde \chi}_{A'B'AC'}\Bigr)
\; \; \; \; .
&(11)\cr}
$$
Note that an analogous result holds if the metric is positive-definite
rather than complex. A four-manifold with nonvanishing torsion and
positive-definite metric is here denoted
by $RU_{4}$. Interestingly, a particular solution of Eq. (11) is given
by ${\widetilde \psi}_{A'B'C'D'}=0$, ${\widetilde \chi}_{A'C'AB'}=0$.
By analogy with complexified general relativity, the particular
$CU_{4}$ and $RU_{4}$ space-times satisfying these additional conditions
are here called right-flat and right-torsion-free (although our
definition does not involve the Ricci tensor, and is therefore different
from Eq. (6.2.1) of Ref. [1]). This means that the
surviving Weyl and torsion spinors, i.e. $\psi_{ABCD}$ and
$\chi_{AB}^{\; \; \; \; \; CC'}$, do not affect the integrability
condition (11) for $\alpha$-surfaces. Note also that this is only possible
for $CU_{4}$ and $RU_{4}$ models of gravity, since only for these theories
the torsion spinors $\chi$ and $\widetilde {\chi}$, and the Weyl spinors
$\psi$ and $\widetilde {\psi}$, are totally independent [1].
\vskip 10cm
\leftline {\bf Acknowledgements}
\vskip 0.3cm
\noindent
We are much indebted to the referees of the journal General Relativity
and Gravitation for suggesting the correct forms of the result and
correcting numerous other errors in the original version of this paper,
and to a referee of Nuovo Cimento for helpful comments which
improved style and nature of our work.
We are also grateful to Professors Giuseppe Marmo and Ruggiero de
Ritis for reading the manuscript and many enlightening conversations,
and to Dr. David Hughes for correspondence on spinor methods.
\vskip 0.3cm
\leftline {\bf References}
\vskip 0.3cm
\item {[1]}
R. S. WARD and R. O. WELLS: {\it Twistor Geometry and Field
Theory} (Cambridge University Press, Cambridge, 1990).
\item {[2]}
G. ESPOSITO: {\it Quantum Gravity, Quantum Cosmology and Lorentzian
Geometries} (Springer-Verlag, Berlin, 1992).
\item {[3]}
R. PENROSE: {\it Found. Phys.}, {\bf 13}, 325 (1983).
\vskip 100cm
\leftline {\bf RIASSUNTO}
\vskip 1cm
\noindent
Questo lavoro studia condizioni necessarie per l'esistenza di
$\alpha$-superfici in modelli di spazio-tempo complesso con torsione non
nulla. Per queste variet\`a, le parentesi di Lie di campi vettoriali e le
identit\`a spinoriali di Ricci contengono esplicitamente gli effetti della
torsione. Questo conduce ad una condizione di integrabilit\`a per le
$\alpha$-superfici che non involve soltanto lo spinore di Weyl auto-duale,
come nella relativit\`a generale complessificata, ma anche lo spinore di
torsione, in modo non lineare, e la sua derivata covariante.
Di un certo interesse,
una soluzione particolare della condizione di integrabilit\`a \`e data da
modelli di spazio-tempo piatti a destra e liberi da torsione a destra.
\bye